\documentclass[a4,rotating,epsfig]{article}
\usepackage{graphicx}
\def\Journal#1#2#3#4{{#1} {\bf #2}, #3 (#4)}

\def\PLB{{\em Phys. Lett.}  {\bf B}}
\def\PRL{\em Phys. Rev. Lett.}
\def\PRD{{\em Phys. Rev.} {\bf D}}

\def\PRP{{\em Phys. Rep. }}
\def\EPC{{\em Eur. Phys. J.} {\bf C}}

\def\MPL{{\em Mod. Phys. Lett. }}

\def\PNPP{\em Prog. Nucl. Part. Phys.}

\def\PTPS{\em Prog. Theo. Phys. Suppl.}

\def\ra{\rightarrow}
\def\be{\begin{equation}}
\def\ee{\end{equation}}

\newcommand{\ls}{\mbox{$\stackrel{<}{\sim}$ }}
\newcommand{\expe}{experiment}

\newcommand{\exps}{experiments}

\newcommand{\br}{branching ratio}
\newcommand{\bb}{double beta decay}

\newcommand{\nbb}{neutrinoless double beta decay}
\newcommand{\majo}{Majorana}

\newcommand{\nel}{\mbox{$\nu_e$}}
\newcommand{\nmu}{\mbox{$\nu_\mu$}}

\newcommand{\neu}{neutrino}
\newcommand{\neus}{neutrinos}

\newcommand{\ema}{\mbox{$\langle m_{ee} \rangle$ }}
\newcommand{\emm}{\mbox{$\langle m_{\mu \mu} \rangle$ }}

\newcommand{\gess}{\mbox{$^{76}Ge$ }}

\newcommand{\Slash}[1]{\mbox{$#1\hspace{-.6em}/$}}

\newcommand{\ba}{\begin{array}{c}}
\newcommand{\baz}{\begin{array}{cc}}
\newcommand{\bad}{\begin{array}{ccc}}
\newcommand{\bea}{\begin{equation} \begin{array}{c}}
\newcommand{\eea}{ \end{array} \end{equation}}
\newcommand{\ea}{\end{array}}
\newcommand{\D}{\displaystyle}

\newcommand{\aen}{\mbox{$\overline{\nu_e} $}}

\newcommand{\mab}{\mbox{$\langle m_{\alpha \beta} \rangle $}}
\newcommand{\mmm}{\mbox{$\langle m_{\mu \mu} \rangle $}}
\newcommand{\mtt}{\mbox{$\langle m_{\tau \tau} \rangle $}}
\newcommand{\mmt}{\mbox{$\langle m_{\mu \tau} \rangle $}}
\newcommand{\mee}{\mbox{$\langle m_{ee} \rangle $}}
\newcommand{\met}{\mbox{$\langle m_{e \tau} \rangle $}}
\newcommand{\mem}{\mbox{$\langle m_{e \mu} \rangle$}}

\begin{document}
\hfill{Contributed paper to Neutrino 2000, Sudbury, 16-21 June 2000}
\vspace{1.5cm}
\begin{center}
{\large \bf {EFFECTIVE MAJORANA NEUTRINO MASSES\footnote{most of the
work done in collaboration with M. Flanz and W. Rodejohann}}}
\vspace{1cm}\\
K. ZUBER\footnote{e-mail: zuber@physik.uni-dortmund.de}
\medskip\\
{\it \normalsize Lehrstuhl f\"ur Experimentelle Physik IV,}\\
{\it \normalsize Universit\"at Dortmund, Otto--Hahn Str. 4,}\\ 
{\it \normalsize 44221 Dortmund, Germany}
\end{center}
\vspace{1cm}
\begin{abstract}
A generalisation of the neutrinoless double beta decay process is presented.
Neutrinoless double beta decay measures only one out of 
nine possible effective Majorana neutrino masses 
in case of three flavours. Limits obtained 
for all the matrix elements - some of them for the first time - 
are presented using data from $\mu - e$ conversion, neutrino-nucleon scattering, 
HERA and rare kaon decays.
An outlook towards future possibilities to improve on the bounds is given.
\end{abstract}

\section{Introduction}
Investigation of lepton--number violating processes is one of the
most promising ways of probing physics beyond the standard model. 
A particular aspect of this topic is lepton--number violation 
in the neutrino sector, which in the case of massive neutrinos
would allow a variety of new phenomena. For recent reviews see \cite{zub98}. 

\begin{figure}
\centering
\includegraphics[width=5cm,height=5cm]{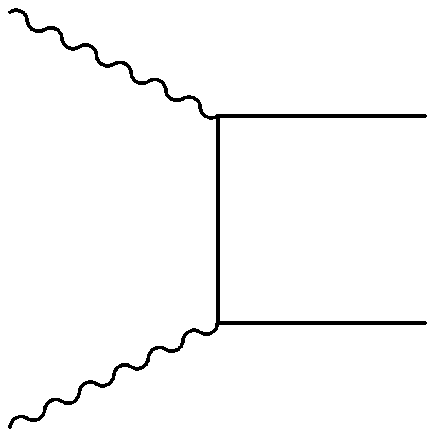} 
\includegraphics[width=5cm,height=5cm]{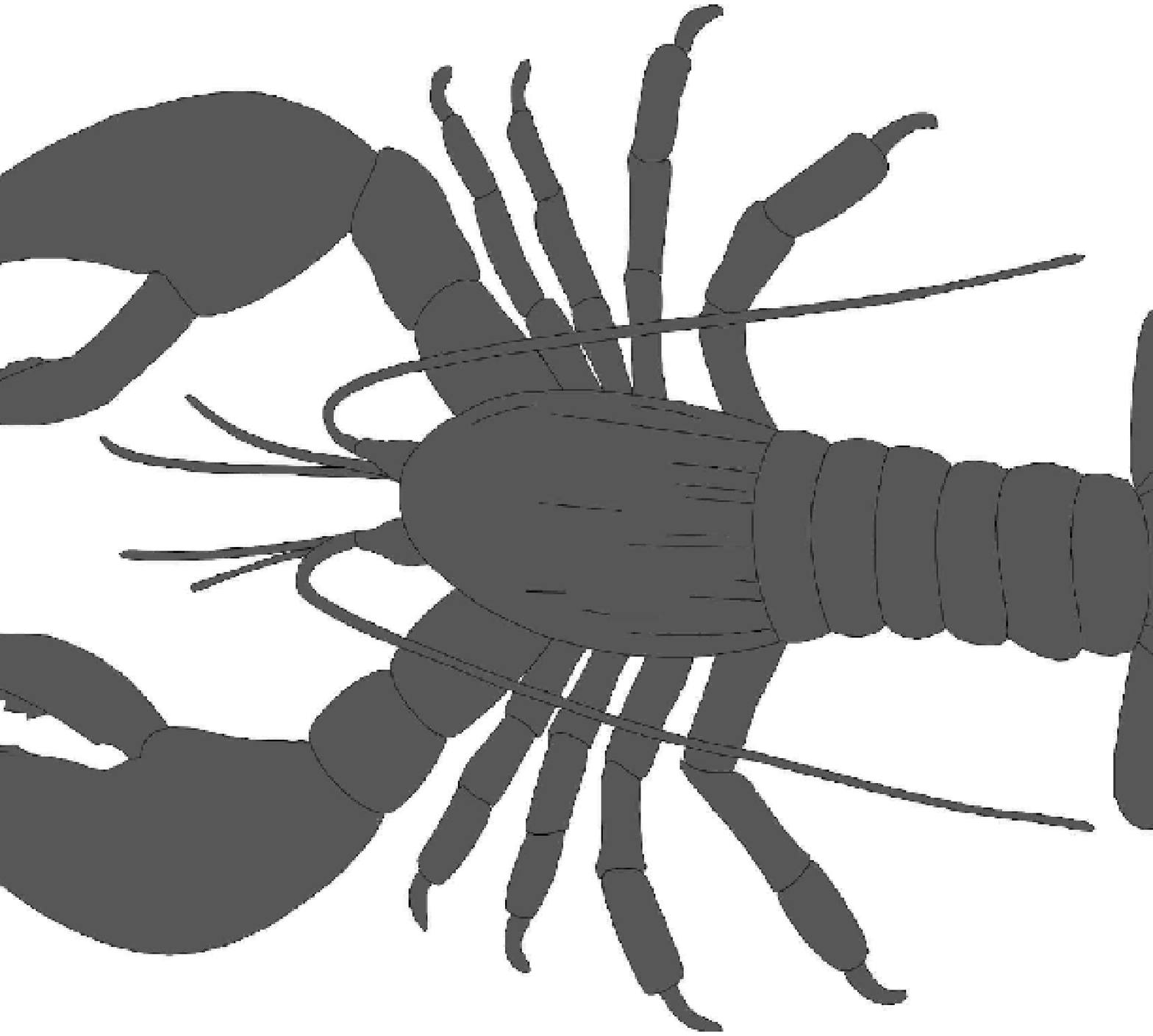}
\caption{\label{pic:feyn}Fundamental Feynman diagram for the processes here
(The ''Lobster'' - diagram).}
\end{figure}

Such processes emerge immediately in case
of Majorana masses of the neutrinos, which are predicted in 
most GUT--theories. Reactions associated with \majo{}
\neus{} are typically characterised by violating lepton number by
two units ($\Delta L$ = 2).\\
The gold plated channel to search for massive Majorana neutrinos 
is \nbb{} of nuclei ($\Delta L_e$ = 2)
\be
(A,Z) \ra (A,Z+2) + 2 e^-
\ee
A lot of activity was spent over the past decades
to push half-life limits for this decay beyond $10^{25}$ yrs and therefore 
the measuring quantity, called effective Majorana mass \ema down below 
1 eV. The best limit currently available is obtained for \gess and results in an upper
limit of \ema of about 0.2 eV \cite{hdmo}. The quantity \ema is given by
\be 
\langle m_{ee} \rangle = \mid \sum U_{em}^2 m_m \eta^{\rm CP}_m \mid 
\ee 
where $m_m$ are the mass eigenvalues, $\eta^{\rm CP}_m = \pm 1$ 
the relative CP--phases and $U_{em}$ the mixing matrix
elements. \\
In general, there is a $3 \times 3$ matrix of effective Majorana masses, 
the elements being 
\bea \label{meffmatrix}
  \mab = |(U \, {\rm diag}(m_1 \eta^{\rm CP}_1
   , m_2 \eta^{\rm CP}_2 , m_3\eta^{\rm CP}_3) 
   U^{\rm T})_{\alpha \beta}| \\[0.3cm]
    = \left| \D \sum m_m \eta^{\rm CP}_m U_{\alpha m} U_{\beta m} \right|
   \mbox{ with }  \alpha, \beta = e , \, \mu  , \, \tau .     
\eea
In contrast to \nbb{} little is known about the other matrix elements. In this 
paper the current status of the full matrix is reviewed as well as suggestions for
future improvements are given.\\
The basic Feynman-graph under investigation is shown in Fig. \ref{pic:feyn}.

\section{$\mu - e$ conversion and \mem }
Muon - positron conversion on nuclei
\be
\mu^- + (A,Z) \ra e^+ + (A,Z-2)
\ee
is a process closely related to double beta decay
and, within the context discussed here, is measuring \mem.
The current best bound is coming from SINDRUMII and is given
by \cite{kau98}
\be
\label{eq:3}
\frac{\D \Gamma ({\rm Ti} + \mu^- \to  {\rm Ca}^{GS} + e^+)}
{\D \Gamma ({\rm Ti} + \mu^- \ra Sc + \nmu)} 
< 1.7 \cdot 10^{-12} \quad (90 \% \mbox{CL}) 
\ee
which can be converted in a new limit of \mem $<$ 17 (82) MeV depending on whether
the proton pairs in the final state are in a spin singlet or triplet state.  
Correction factors of the order one for the difference in Ti and S as 
given in \cite{doi85} might be applied.
As can be seen this limit is already about 8 orders of magnitude worse than \nbb.
Improvements on \mem{} by an order of magnitude can be expected by a new run of SINDRUM II
in 1999. A further step forward might be the proposed AGS- E940 \expe{} (MECO) at BNL and PRISM at the 
Japanese Hadron Facility.\\
Notice that a process like $\mu \ra e \gamma$ does not give direct bounds on the quantities
discussed here, because it measures $m_{e\mu}= \sqrt{\sum U_{ei} U_{\mu i} m_i^2}$.
Therefore without specifying a \neu{} mixing and mass scheme, the quantities are rather difficult 
to compare. On the other hand, if this is done, such an indirect bound is more stringent.  

\section{Trimuon production in $\nu N$ - scattering and \mmm}
\begin{figure}[hp]
\setlength{\unitlength}{1cm}
\begin{center}
\includegraphics[width=13cm,height=8cm]{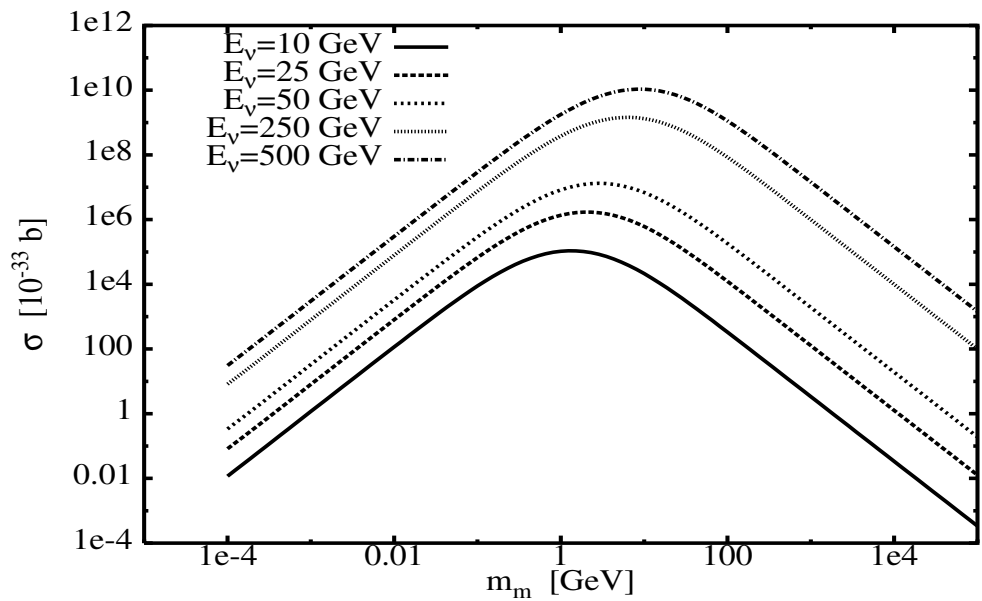}
\end{center}
\vspace{-0.7cm}
\caption{\label{pic:trim}Total cross section of the process $\nu_{\mu} N \ra
\mu^- \mu^+ \mu^+ X$ for a left--handed
Majorana neutrino as a function of its mass for different neutrino beam 
energies. No limit on $U_{\mu m}^{2}$ was applied. The obtained cross sections
are tiny.}
\end{figure}

The process under study is muon lepton--number violating 
($\Delta L_{\mu} = 2$) trimuon production in neutrino--nucleon scattering via charged 
current reactions (CC)
\be
\label{proces}
\nmu N \ra \mu^- \mu^+ \mu^+ X 
\ee
where $X$ denotes the hadronic final state.
The measured mixing matrix element is \mmm{}. Detailed calculations can be found in \cite{frz98}. 
Taking the fact that in past \exps{} no excess events of this type were observed
on the level of $10^{-5}$ with respect to \nmu CC,
a limit of $\mmm \ls 10^4$ GeV\@ can be deduced. This has to be compared to \emm $\! < 1.1 \cdot 10^5$
GeV
as obtained from earlier K--decay data \cite{nis99}. Even being an order of magnitude improvement
the obtained bound is -- like the kaon bound -- still in an unphysical region 
because it is a simple extrapolation from small neutrino masses upwards neglecting effects 
from the propagator term. This means that the real cross section scales with
\be
  \sigma \propto
  \left| 
  \sum\limits_m \frac{m_m \eta_m^{\rm CP}  U_{\mu m}^2 }
  {(q_2^2 - m_m^2)}
  \right|^2 .
\ee
which leads to a $m_m^{-2}$ behaviour in case that $m_m^2 >> q^2$. 
The total cross section for various \neu{} energies is shown in Fig. \ref{pic:trim}.\\
Current \exps{} like CHORUS and NOMAD might improve on this quantity by an order of magnitude. 
\section{Rare kaon decays and \mmm}
As already mentioned, a further possibility to probe \mmm{} is the rare kaon decay 
\be
K^+ \ra \pi^- \mu^+ \mu^+ \quad .
\ee
Detailed calculations can be found in \cite{hal76,lit92}.
Using old data a \br{} of
\be
\frac{\D \Gamma (K^+\to \pi^- \mu^+ \mu^+)}{\D \Gamma (K^+ \to {\rm all})} 
< 1.5 \cdot 10^{-4} \quad (90 \% \mbox{CL})
\ee
was obtained \cite{lit92}. 
Combined with the theoretical calculations of \cite{doi85} a limit of \mmm $< 1.1 \cdot 10^5$
GeV could
be deduced \cite{nis99}.
In the meantime new sensitive kaon experiments are online and using the E865 \expe{} at BNL
a new upper limit on the \br{} of 
\be
\frac{\D \Gamma (K^+ \to \pi^-\mu^+ \mu^+)}{\D \Gamma (K^+ \to {\rm all})} 
< 3 \cdot 10^{-9} \quad (90 \% \mbox{CL})
\ee
could be deduced \cite{ma99}, an improvement by a factor 50000. Assuming the \br{} scales with $\mmm^2$
(valid as long as $m_m \ll q^2$) this can be converted in the limit \mmm \ls 500 GeV \cite{zub00}, a factor
of eight better
than the existing limits coming from HERA (see sec. \ref{s5}) and three orders of magnitude
better for this particular decay channel.
Also here the propagator term is neglected and therefore 
even this improvement still does not imply any serious limit on heavy neutrinos
\cite{lit00}.
To improve on \mmm{} the decay of charmed mesons could be considered as
well. Among the Cabibbo favoured modes are 
$D^+ \ra \pi^- \mu^+ \mu^+, D_S^+ \ra K^- \mu^+ \mu^+$ or $D_S^+ \ra \pi^- \mu^+ \mu^+$. 
The existing limits on the \br{} for these processes are
$1.7 \cdot 10^{-5}, 1.9 \cdot 10^{-4}$ and $8.2 \cdot 10^{-5}$
respectively \cite{ait99}. While being competitive with the old bound for the kaon decay discussed, the
new kaon \br{} limit is now four orders of magnitude better.
Therefore, to obtain new information on \mmm{} from D$^+$-decays, analyses of new data sets have to be done.\\
To improve significantly towards lighter \neu{} masses (\mmm \ls 1 GeV) 
one might consider other processes. 
The close analogon to \bb{} and also a measurement of \mmm{} using nuclear scales would be
$\mu^-$ -- capture by nuclei with a $\mu^+$ in
the final state
as discussed in \cite{mis94}. No such \expe{} was performed yet, probably because of the requirement to use 
radioactive targets due to energy conservation arguments.
The ratio with respect to standard muon capture can be written in case of the favoured
$^{44}$Ti and light neutrino
exchange 
($m_m \ll q^2$) as
\be
  R = \frac{\Gamma (\mu^- + {\rm Ti} \ra \mu^+ + {\rm Ca})}{\Gamma (\mu^- + {\rm Ti} \ra \nmu + {\rm Sc})}
  \simeq 5 \cdot 10^{-24} (\frac{\mmm}{250 keV})^2 \quad .
\ee 
Assuming that a \br{} of the order of muon-positron
conversion (eq.\ref{eq:3}) can be obtained, a bound \mmm \ls 150 GeV results. Unfortunately
this is only a slight improvement on the bound obtained from kaon decay and to make real progress
a branching ratio of the order $R \approx 10^{-17}$ has to be measured.
Assuming heavy neutrino exchange ($m_m \gg q^2$) for the muon capture,
would result in a rate another four orders of magnitude lower than for light neutrino exchange.

\section{Limits from HERA on the full mass matrix}
\label{s5}
A first set of full matrix elements including the $\tau$ - sector was given by \cite{frz} using HERA data.
The process studied is

\be  \label{process}
  e^{\pm} p \ra \stackrel{(-)}{\nel}
  l^{\pm} l'^{\pm} X, \mbox{ with } (l l') = (e \tau) , \; 
  (\mu  \tau) , \;(\mu  \mu) \mbox{ and } (\tau \tau) 
\ee

\begin{figure}[hb]
\setlength{\unitlength}{1cm}
\vspace{0.3cm}
\begin{center}
\includegraphics[width=13cm,height=8cm]{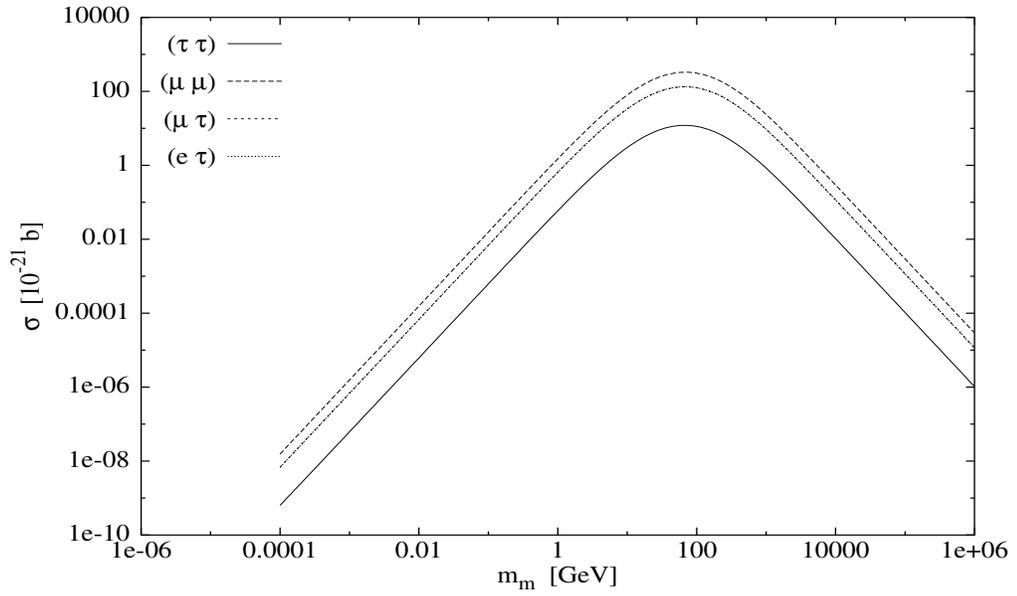}
\end{center}   
\caption{\label{pic:hera}Total cross section for the process
$e^{+} p \ra \aen l^{+} l'^{+} X$ at HERA as a function of {\it one} eigenvalue
$m_m$.
No limits on $U_{l m}$ are applied, the branching ratio for taus into muons is
included. The $(e \tau)$ and
$(\mu \tau)$ cases are indistinguishable in this plot.}
\end{figure}

Such a process has a spectacular signature with 
large missing transverse momentum ($\Slash{p}_T$) and two like--sign 
leptons, isolated from the hadronic remnants.
The mass bounds given below are obtained for analysed luminosities
of $\mbox{${\cal L}$}_{e^+} = 36.5  $ pb$^{-1}$ (H1) and
$\mbox{${\cal L}$}_{e^+} = 47.7  $ pb$^{-1}$ (ZEUS) and are typically in
the range $10^3 - 10^4$ GeV and given in eq. \ref{meffresult}.
The cross section is shown in Fig. \ref{pic:hera} including indirect
bounds on mixing elements deduced from other \exps{} \cite{nar95}.\\
An extension of the analysis allowing for any two final
state leptons and the possibility of observing only one isolated lepton
(because of applied cuts) is given in \cite{rz00}. This is studied within
the context of the published unusual large number of events with single isolated leptons
and large $\Slash{p}_T$ observed by H1 \cite{adl98}.\\ 
Possible upgrades of HERA and a more sophisticated general analysis of $\tau$ - decays
might allow an improvement by a factor 20.\\
As stated before, applying limits from flavour changing neutral currents like
$\tau \ra \mu \gamma$ might be more stringent but require the specification
of a mixing and mass scheme.

\section{Future prospects}
Beside the already mentioned steps for improvements two ways might be worthwile to follow.
First of all more general meson decays. Improvements on the $\tau$ - sector of matrix elements,
especially \mtt, could be done by
a search for rare B-decays. Limits on the \br{} for decays $B^+ \ra K^- \mu^+ \mu^+, B^+ \ra \pi^-
\mu^+ \mu^+$ of less than $9.1 \cdot 10^{-3}$ exist \cite{wei90}, however nobody looked into the decays $B^+
\ra K^- \tau^+
\tau^+$ or $B^+ \ra \pi^- \tau^+ \tau^+$. With the new B-factories such a search might be possible
at a level of producing limits on \mtt{} competitive with the ones given. \\
A significant step forward is possible by using the proposed neutrino factory. Such a high luminosity
neutrino machine, producing $10^{13}$ charged current interactions per year in a near detector, allow to
improve significantly on the trimuon production process as discussed in \cite{rz001}. Furthermore in
a ''parasitic'' mode also improved searches for rare kaon decays are foreseen, which should include
$K^+ \ra \pi^- \mu^+ \mu^+$. This might allow to bring \mmm{} in a physical useful region.

\section{Summary and conclusions}

The underlying physical process of \nbb{} by exchange of a virtual massive
\majo{} \neu{} state was extended to three flavours. Processes studied
within that context are reviewed.\\
Combining all obtained limits, ignoring possible phases in the elements   
$U_{\alpha m}$ (therefore getting a symmetrical matrix $\mab{}$) 
as well as skipping the intrinsic CP parities,
the following bounds for the effective Majorana mass matrix emerge: 

\be \label{meffresult}
    \mab = \left( \bad \mee & \mem & \met \\[0.2cm]
                        & \mmm & \mmt \\[0.2cm]
                        &      & \mtt \ea \right) \ls 
    \left( \bad 2 \cdot 10^{-10} & 1.7 (8.2) \cdot 10^{-2} & 4.2 \cdot 10^{3}\\[0.2cm]
                              &  500 & 4.4 \cdot 10^{3} \\[0.2cm]
                              &           & 2.0 \cdot 10^{4} 
    \ea \right) \rm GeV .  
\ee

A spread over 14 orders of magnitude can be seen. Again, 
these are direct limits and all elements other than \mee{} and \mem{} are
still in an unphysical region because propagator effects are ignored. To bring
at least \mmm{} in the physical region as well, the \neu{} factory could help a lot
because improving both on trimuon production and rare kaon decays described above. 
This would not only allow to put upper limits on light masses, but also lower limits
on heavy neutrinos can then be obtained.

\newpage


\end{document}